\documentclass[12pt,preprint]{aastex}               
\shortauthors{D.~A.~Uzdensky}

\begin{document}

\title{Petschek-like reconnection with current-driven anomalous 
resistivity and its application to solar flares}
\author{Dmitri A. Uzdensky}
\affil{Kavli Institute for Theoretical Physics, University of California} 
\affil{Santa Barbara, CA 93106}
\email{uzdensky@kitp.ucsb.edu}
\date{\today}

\begin{abstract}
Recent numerical simulations of magnetic reconnection in two 
dimensions have shown that, when the resistivity is strongly 
localized, the reconnection region develops a Petschek-like 
structure, with the width of the inner diffusion region 
being of the order of the resistivity localization scale. 
In this paper, we combine this fact with a realistic model 
for locally-enhanced anomalous resistivity generated by 
current-driven microturbulence. The result is a qualitative 
model of the reconnection layer where the size of Petschek's 
diffusion region and, therefore, the final reconnection rate
are determined self-consistently in terms of the main parameters 
of the functional dependence of anomalous resistivity on the 
electric current density. We then consider anomalous resistivity 
due to ion-acoustic turbulence as a particular case. This enables 
us to express the reconnection region's parameters directly in
terms of the basic parameters of the plasma. Finally, we apply 
this reconnection model to solar flares and obtain specific 
predictions for typical reconnection times, which are very
consistent with observations.  
\end{abstract}

\keywords{MHD --- Sun: flares --- Sun: magnetic fields}


\section{Introduction}
\label{sec-intro}

Magnetic reconnection is a basic plasma physics phenomenon
of tremendous importance in many astrophysical systems
(Tsuneta~1996; Kulsrud~1998), as well as in some laboratory 
plasma devices (Yamada~et~al. 1997), including tokamaks 
(Kadomtsev~1975; Yamada~et~al. 1994). It has been studied 
extensively over the past five decades (e.g., Giovannelli~1946;
Vasyliunas~1975; Biskamp~2000). Historically, the first (and 
perhaps still the most important) application of magnetic 
reconnection has been to explain the solar flare phenomenon, 
and it is in this context that the earliest reconnection models 
have been developed.

The first theoretical model of magnetic reconnection was 
developed by Sweet (1958) and by Parker (1957, 1963). In 
this model, magnetic field is frozen into the plasma 
everywhere except in a very thin layer where the 
current density is so high that resistive effects become 
important no matter how small the resistivity is. It is 
inside this thin current layer that the actual breaking (and reconnecting) 
of the magnetic lines of force takes place, accompanied by 
a violent release of enormous amounts of magnetically stored 
energy, thus leading to the observed flare. The Sweet--Parker 
theory predicts that the current layer thickness, $\delta_{\rm SP}$, 
scale as $\delta_{\rm SP} \sim L/\sqrt{S}$, where $L$ is the 
global system size (typically of order $10^9$ cm in the solar 
corona) and $S\equiv L V_A/\eta$ is the global Lundquist 
number (here $V_A$ is the Alfv{\'e}n velocity and~$\eta$
is the magnetic diffusivity; in the rest of this paper we
shall refer to $\eta$ as the resistivity). Correspondingly, 
the typical reconnection timescale is found to be 
$\tau_{\rm rec}\sim \tau_A(L) \sqrt{S}$, where $\tau_A(L)\equiv L/V_A$
is the global Alfv{\'e}n transit time. It has been immediately
realized that the resulting reconnection time turns out to be
too long; in the solar corona one typically has $S\sim 
10^{12}-10^{14}$ and $\tau_A(L)\sim 1$~sec, which leads
to $\tau_{\rm rec}$ of the order of a few months. This is
in sharp contrast with the typical observed solar flare 
duration of order $10^2 - 10^3$~sec. Thus, since its early 
years, the main thrust of magnetic reconnection research has 
been to explain reconnection rates that are much faster than 
the Sweet--Parker theory predicts.

Starting from the 1960s, two major routes toward faster reconnection
were proposed. One of them was to use the so-called {\it anomalous 
resistivity} instead of the classical Spitzer resistivity used in 
the original Sweet--Parker model (Coppi \& Friedland 1971; Smith 
\& Priest 1972; Coroniti \& Eviatar 1977; Kulsrud~1998). The idea 
was that, as a current layer forms, its thickness becomes so small, 
and hence the current density becomes so high, that the drift velocity 
of the current-carrying electrons exceeds a certain threshold, such as 
the ion-sound or the electron thermal speed. This leads to the excitation 
of current-driven kinetic microturbulence, which, in turn, provides a 
more efficient (compared to particle-particle collisions) mechanism for 
the scattering of electrons (via wave-particle interactions). As a result, 
one ends up with a greatly enhanced effective resistivity and, hence, a 
greatly reduced effective Lundquist number. When substituted into the 
Sweet--Parker scaling, this results in a greatly enhanced reconnection 
rate. In fact, controlled laboratory studies of magnetic reconnection 
have shown a good agreement with a simple Sweet--Parker model augmented 
with some (experimentally measured) anomalously enhanced resistivity 
(Ji~et~al. 1998, 1999). On the other hand, in the solar flare context, 
the Sweet--Parker model with anomalous resistivity gives typical 
reconnection times of the order of several hours (e.g., Kulsrud~1998), 
a great improvement over Spitzer resistivity. Among various anomalous 
resistivity mechanisms, the one most frequently quoted has been the 
{\it ion-acoustic turbulence} (IAT). Rapid magnetic energy dissipation 
due to the IAT-driven anomalous resistivity has in fact also been invoked 
to explain coronal heating (Rosner~et~al.~1978).

Another possibility leading to shorter reconnection times was 
proposed by Petschek (1964). His very elegant model makes use 
of a somewhat more complicated reconnection layer geometry, 
while still relying on simple resistive magnetohydrodynamics 
(MHD) without invoking any new physics at the microscopic level. 
The {\it Petschek model} actually does not predict a unique 
configuration of the reconnection layer and a unique reconnection 
rate. Instead, this model encompasses an entire one-parametric 
family of solutions. Each of these configurations has at its 
center a small Sweet--Parker-like layer, called the {\it diffusion 
region}, and four standing slow-mode shocks emanating from the ends 
of this central layer. The members of this family of solutions can 
be labeled by the width (or length) $\Delta$ of the inner diffusion 
region. As Petschek noticed, in the Sweet--Parker model the 
reconnection process had been slowed down by the very large 
aspect ratio $L/\delta$ of the reconnection layer. He suggested 
that the width~$\Delta$ of the layer's diffusion region does not 
have to be as large as the global size~$L$; if $\Delta$ can be 
made sufficiently short, the reconnection process will go much 
faster than in the Sweet--Parker model. The maximum value of~$\Delta$ 
corresponds to the Sweet--Parker solution, which is, therefore, just 
one of the family members. The reconnection rate ranges from the 
slowest (Sweet--Parker) rate to the so-called maximum Petschek rate, 
which scales as $1/\log{S}$. We thus see that the relatively strong
square-root dependence on the resistivity, characteristic for the 
Sweet--Parker model, is replaced here by a much weaker logarithmic
dependence; even for a very large $S\sim 10^{14}$, the resulting
reconnection timescale turns out to be reasonably short, of order 
$10^2\, \tau_A(L)$.

This model had remained the favorite model of reconnection until 1980s, 
when two-dimensional (2D) resistive-MHD numerical simulations by Biskamp 
(1986) showed that, in the case of a spatially uniform resistivity, 
a Petschek-like configuration fails to form and that a long (of 
order~$L$) current layer tends to form instead, consistent with 
the Sweet--Parker picture. This finding has been confirmed in a 
number of numerical simulations performed by several other groups 
(Scholer~1989; Ugai 1992, 1999; Yokoyama \& Shibata 1994; Uzdensky 
\& Kulsrud~2000; Erkaev~et~al. 2000, 2001). A theoretical explanation 
has been put forward by Kulsrud (2001). He noticed that, when the
resistivity is uniform, the relatively large transverse magnetic field
that is needed to support the standing shocks in the Petschek model 
is rapidly swept out of the diffusion region by the downstream flow,
while its regeneration due to the nonuniform merging is not fast 
enough. As a result, the diffusion region's size~$\Delta$ (which 
Kulsrud calls~$L'$) increases until it reaches the global scale~$L$
and the reconnection rate, correspondingly, slows down to the usual
Sweet--Parker rate. This explanation has been confirmed numerically 
by Uzdensky \& Kulsrud (2000) (see also Kulsrud~1998).

It has been noticed, however, that the key assumption leading to 
the above conclusion was the assumption of uniform resistivity 
(which is a very common assumption in numerical simulations in 
general). Simulations featuring a {\it non-uniform resistivity} 
have shown that a Petschek-like structure does form and can be 
stable whenever the resistivity is locally enhanced in some small 
region near the X-point at the center of the reconnection region 
(Ugai \& Tsuda 1977; Sato \& Hayashi 1979; Ugai 1986, 1992, 1999; 
Scholer~1989; Yokoyama \& Shibata 1994; Erkaev et al. 2000, 2001; 
Ugai \& Kondoh 2001; Biskamp \& Schwarz 2001). A nice plausible 
theoretical explanation of this phenomenon has again been provided
by Kulsrud (2001), who analyzed how a locally enhanced resistivity
may lead to a more efficient regeneration of the transverse magnetic
field through non-uniform merging and thus to the sustainment of
Petschek's shocks. In addition, recent numerical work by Erkaev
et~al. (2000, 2001) and by Biskamp \& Schwarz (2001) has shown 
that the particular Petschek-like configuration that forms in 
the localized-resistivity situation is characterized by the 
width~$\Delta$ of the inner diffusion region being of the order 
of the resistivity localization scale, which in this paper we shall 
call~$l_\eta$ (Erkaev~et~al. 2000, 2001; Biskamp \& Schwarz 2001). We 
thus see that resistivity-localization mechanism plays a crucial 
role in determining the final reconnection rate.

From the point of view of physical reality, 
the main motivating force behind these localized-resistivity
studies has been the idea that anomalous resistivity, being such 
a sensitive function of the local current density, may in fact be 
triggered only in a small neighborhood of the X-point, where the 
current density is highest. In other words, anomalous resistivity
can enhance reconnection rate not only directly (by simply being
higher than the collisional resistivity), but also indirectly, 
via enabling the Petschek mechanism (by being strongly localized).
This is one of the key ideas behind the so-called spontaneous fast
reconnection model suggested by Ugai (1986, 1992, 1999; also see
Ugai \& Tsuda 1977; Yokoyama \& Shibata 1994; Ugai \& Kondoh 2001)
on the basis of numerical simulations.

Up until now, however, with the notable exception of the work
by Kulsrud (2001), there have been no analytical attempts to 
combine the Petschek reconnection model with any physically 
realistic model of anomalous resistivity, which would provide 
unique theoretical predictions for the main parameters of the 
reconnection region, such as the width of the diffusion region
and the reconnection rate, in terms of the basic parameters of 
the plasma. The goal of this paper is to remedy this situation 
by attempting to build a simple theoretical framework incorporating 
a particular anomalous resistivity mechanism (ion-acoustic turbulence).

In \S~\ref{sec-model} we present our model of the Petschek
reconnection layer with a generic form of anomalous resistivity.
In particular, in \S~\ref{subsec-ingredients} we describe
the basic elements of the model, e.g., the Sweet--Parker 
relationships for the inner diffusion region and the functional
dependence of anomalous resistivity, $\eta$, on the current
density~$j$. In \S~\ref{subsec-stability} we discuss possible
solutions and explore their stability. In \S~\ref{subsec-rate}
we calculate the reconnection rate in terms of the model parameters.
\S~\ref{sec-IAT} is devoted to the specific case of anomalous resistivity
due to the IAT; in this section we make use of the well-developed
theory of ion-acoustic turbulence to express all major reconnection
layer parameters, including the reconnection rate, in terms of the
basic plasma parameters (such as the magnetic field strength, plasma
density, and the electron and ion temperatures). We apply the obtained
results to solar flare environment in~\S~\ref{sec-flares} and find
a very reasonable agreement in terms of general timescales. We list
our conclusions in~\S~\ref{sec-conclusions}.


\section{Model of the Reconnection Layer}
\label{sec-model}

\subsection{Three Main Ingredients of the Model}
\label{subsec-ingredients}

We now describe our model of a Petschek-like reconnection 
configuration that is formed in the presence of anomalous 
resistivity due to a current-driven microturbulence. This 
model is semi-empirical and does not pretend to describe 
any real physical system in full detail. We believe, however, 
that it correctly captures the most critical qualitative 
features of the reconnection phenomenon.

Our model represents a synthesis of the following three 
external ingredients that are combined to build a 
complete and, hopefully, self-consistent description of 
the system: \\
1) the numerically-observed fact that, whenever the resistivity is
strongly localized, a Petschek-like structure tends to develop, with 
a characteristic width of the central diffusion region being of the 
order of the resistivity localization scale (Erkaev~et~al. 2000, 2001; 
Biskmap \& Schwarz 2001); \\
2) Sweet--Parker model for the central diffusion region of the 
Petschek configuration; \\
3) a physically motivated model for the anomalous resistivity.

\begin{figure}
\plotone{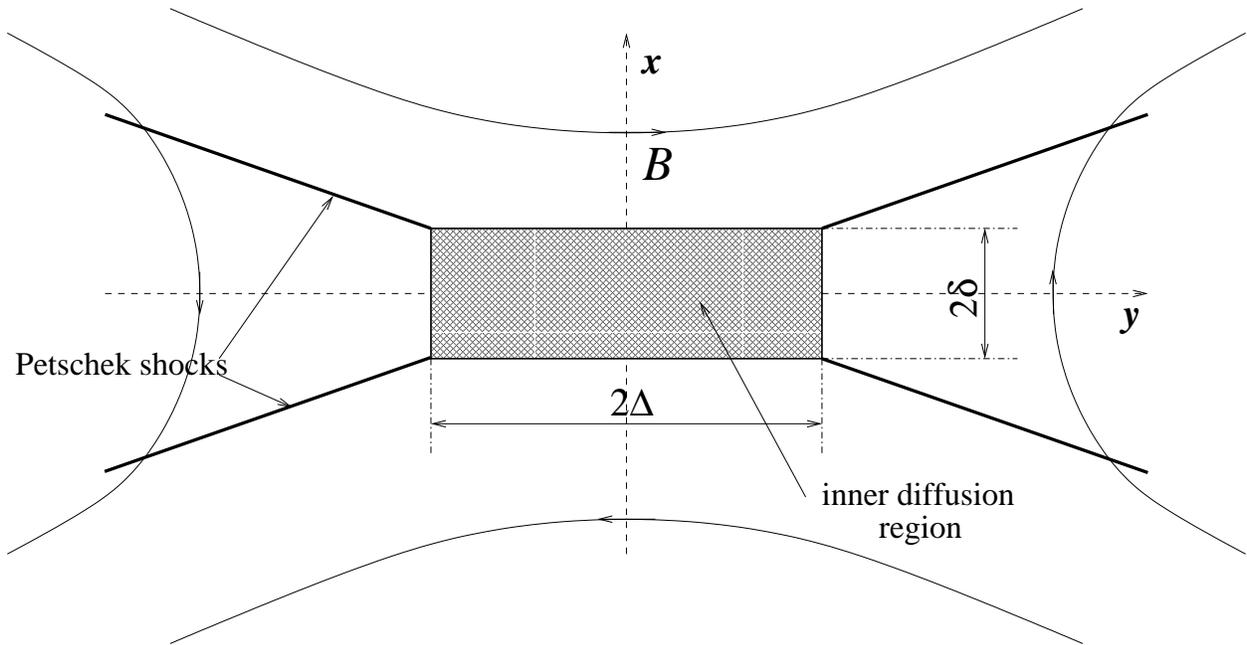}
\figcaption{A schematic drawing of the inner diffusion region 
in the Petschek model.
\label{fig-geometry}}
\end{figure}

We shall now proceed with the description and integration of these
three main ingredients. Let $y$ be the direction along the layer 
and~$x$ across the layer. Figure~\ref{fig-geometry} shows schematically 
the central part of the Petschek-like configuration discussed in this 
paper. The shaded rectangular area (of characteristic thickness~$\delta$ 
and width~$\Delta$) in the center of the Figure is the inner diffusion 
region where the electric current density is concentrated. This is the 
region where the effective plasma resistivity plays an important role 
and where the actual breaking of magnetic field lines takes place. 

For definiteness, we use Ampere's law to define the thickness $\delta$ 
of this central current layer in terms of the central current density 
$j_0\equiv j(0,0)$ and the outside magnetic field $B_0$ as
\begin{equation}
\delta \equiv {{cB_0}\over{4\pi j_0}}\, .
\label{eq-def-delta}
\end{equation}

In addition, suppose that along the midplane $x=0$ the current density 
has a certain profile~$j(y)$. 
We define the characteristic width~$\Delta$ 
of the diffusion region in terms of the function~$j(y)$ as the distance 
from the center $y=0$ to the point where the current density drops by a 
factor of~$e$:
\begin{equation}
j(y=\Delta) = j_0/e \, .
\label{eq-def-Delta}
\end{equation}

Similarly, we define the characteristic scale~$l_\eta$ for the variation of 
the plasma resistivity $\eta(y)$ along the layer as 
\begin{equation}
\eta(y=l_\eta)=\eta(y=0)/e \, .
\label{eq-def-l_eta}
\end{equation}

In general, the two scales $\Delta$ and~$l_\eta$ may be very different.
However, as discussed in the introduction, recent numerical simulations 
have shown that, if the resistivity is strongly localized ($l_\eta\ll L$, 
where $L$ is the global system size), a Petschek-like configuration
develops with the width of the inner diffusion region 
being of order of the resistivity localization scale:
$\Delta \sim l_\eta$. This important condition
serves as the criterion for selecting a unique solution 
out of the entire family of Petschek configurations. 
We shall call this unique Petschek configuration an
equilibrium configuration. 
In our present analysis we shall rely on 
this empirical finding as on one of the important building blocks of 
our model.

Note that, in general, $\Delta$ and $l_\eta$ in this equilibrium
Petschek configuration may differ by a finite factor, so let 
us define a dimensionless parameter~$K$ such that the equilibrium 
configuration has $\Delta =\Delta_{\rm eq}\equiv Kl_\eta$. In our 
analysis it will actually be more convenient to use another 
dimensionless parameter to describe the equilibrium Petschek 
configuration. We shall call this parameter~$\xi$ and define 
it in terms of $j(y)$ as $j(l_\eta)=\xi j_0/e$. The parameters~$K$ 
and~$\xi$ are related to each other, but the exact relationship 
depends on the detailed profiles of $\eta(y)$ and~$j(y)$; it is, 
however, unimportant for the purposes of the present paper: we 
shall only use the fact that, since $j(y)$ is a monotonically 
decreasing function of~$y$, we have $\xi>1$ whenever $\Delta_{\rm eq}
>l_\eta$ ($K>1$) and $\xi<1$ whenever $\Delta_{\rm eq}<l_\eta$ ($K<1$). We 
expect $\xi$ to be of order one, its precise value depends on the 
details of the problem and must be determined from numerical 
simulations. The analysis in this paper will be restricted to 
a situation where $\Delta_{\rm eq}$, i.e., the equilibrium value 
of~$\Delta$, is somewhat greater than $l_\eta$; thus we take $\xi>1$. 
This choice is made purely for reasons of convenience, as will be 
elucidated below.

We shall now describe the other two main components of the model.
The second ingredient is the model for the inner diffusion region. 
In Petschek's theory, this region is a Sweet--Parker-like current 
layer with the thickness~$\delta$ related to its width~$\Delta$ via 
the relationship:
\begin{equation}
{\delta\over\Delta} \simeq S_{\Delta}^{-1/2}\, ,
\label{eq-SP-scaling}
\end{equation}
where
\begin{equation}
S_{\Delta} \equiv {{V_A \Delta}\over{\eta_{\rm eff}}}
\label{eq-def-S_Delta}
\end{equation}
is the Lundquist number for the scale~$\Delta$.
Here we shall take $\eta_{\rm eff}=\eta(j_0)$, the resistivity
at the very center of the layer. Then we can rewrite the above 
expression for~$\delta$ as
\begin{equation}
\delta = \sqrt{{\Delta\over V_A} \eta(j_0)} \, .
\label{eq-SP-delta}
\end{equation}
This is a very important relationship that we are going to invoke 
many times throughout the paper.

The next question is what actually determines the value of~$l_\eta$ and 
hence~$\Delta$. We suggest that these values have to be determined 
by the properties of the function~$\eta(j)$. We thus need to introduce 
the third ingredient, namely, a physically plausible model for anomalous 
resistivity, expressed in terms of a single function~$\eta(j)$. 

Our choice of the function $\eta(j)$ is motivated by anomalous 
resistivity due to a current-driven microturbulence. Correspondingly, 
we here adopt a very simple, minimal model for $\eta(j)$, which, 
however, has to exhibit the following general properties: first, 
there exists a current-density threshold, $j_c$, for triggering 
anomalous resistivity, and second, the rapid rise of~$\eta$ after 
the threshold is exceeded stops at some large but finite value~$\eta_1$, 
after which $\eta(j)$ continues to rise with increased~$j$ more slowly.

\begin{figure}
\plotone{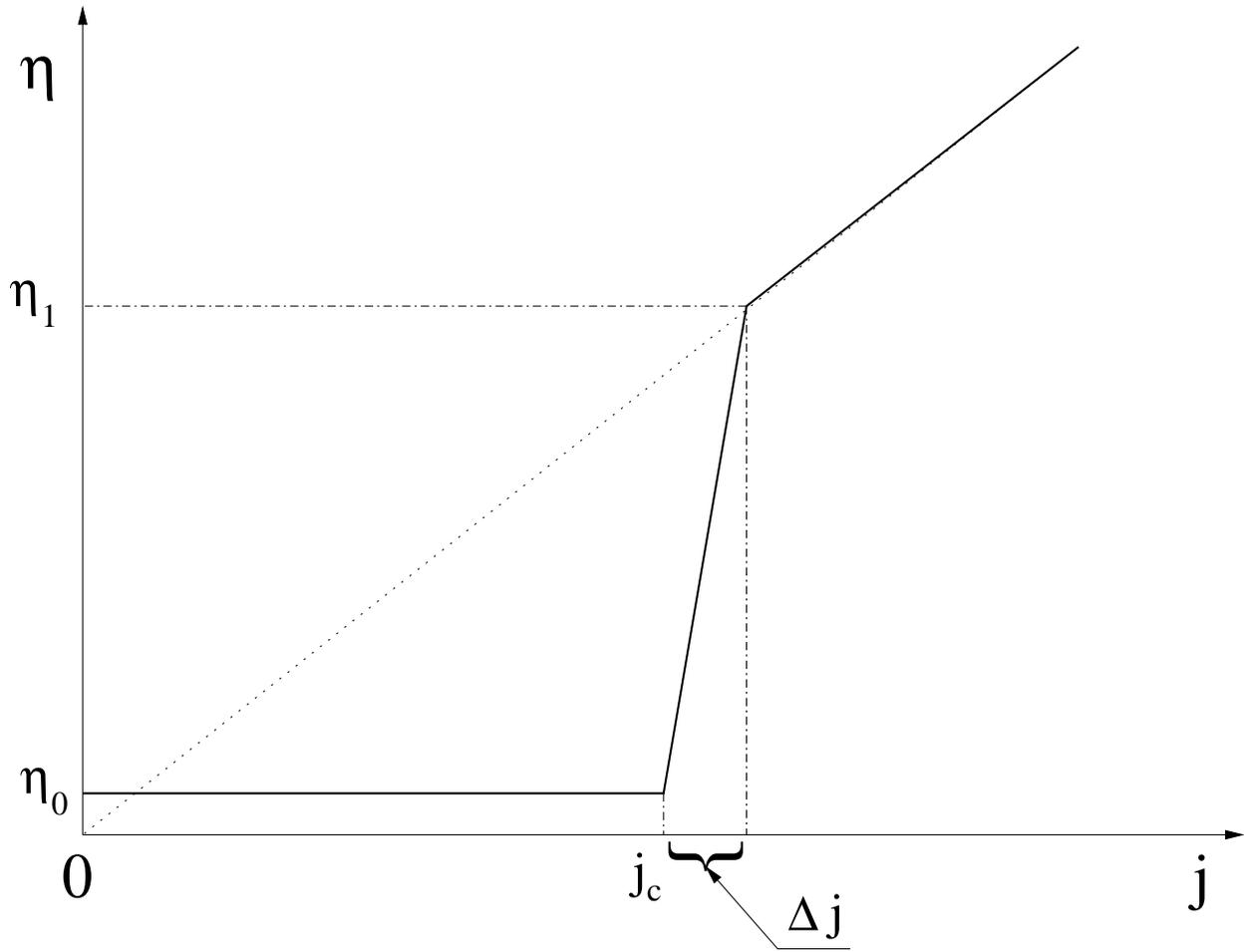}
\figcaption{The anomalous resistivity model adopted in this paper.
\label{fig-resistivity}}
\end{figure}

Thus we take $\eta(j)$ to be a prescribed function whose behavior 
can be described as follows (see Fig.~\ref{fig-resistivity}): \\
1) For $j<j_c$, $\eta(j)$ is constant and equal to a 
small collisional resistivity~$\eta_0$. \\
2) At the critical current density, $j=j_c$,
$\eta$ rises rapidly, essentially jumps to a value $\eta_1\gg \eta_0$.
For our convenience, we actually introduce some small finite width
to this jump, $\Delta j \ll j_c$, and assume that $\eta(j)$ is
a linear function in the interval $j\in[j_c, j_c+\Delta j]$.
The exact value of $\Delta j$ is unimportant in our analysis. \\
3) When $j$ is increased even further, $j>j_c+\Delta j$, 
the resistivity continues to rise monotonically with 
increased $j$, but in a much slower manner.%
\footnote
{Numerical simulations sometimes assume a stronger saturation
of anomalous resistivity by imposing a strict upper limit 
on~$\eta$ (Yokoyama \& Shibata 1994).}
In particular, we take $\eta\sim j$ in this region, which is the 
case for ion-acoustic turbulence (e.g., Bychenkov~et~al. 1988). 
Thus we take
\begin{equation}
\eta(j>j_c+\Delta j) \simeq \eta_1 {j\over{j_c}} \, .
\label{eq-def-eta1}
\end{equation}

We see that the dependence of the anomalous resistivity~$\eta$ 
on the current density~$j$ involves three parameters, $j_c$, 
$\eta_0$, and~$\eta_1$. In addition to these parameters, it is convenient to 
introduce the critical layer thickness $\delta_c$ as a 
derived quantity:
\begin{equation}
\delta_c \equiv {{cB_0}\over{4\pi j_c}}\, .
\label{eq-def-delta_c}
\end{equation}

For the analysis in this section this generic level of description
will suffice. In \S~\ref{sec-IAT}, however, we shall consider a very 
important specific example of anomalous resistivity due to ion-acoustic 
turbulence and will give specific expressions for these parameters.

Here, however, we wish to make a remark concerning the manner 
in which we are going to use the above prescription for~$\eta(j)$.
In our model, $\eta$ is determined solely by the local value of~$j$. 
We realize, of course, that in reality the coefficients~$j_c$, 
$\eta_0$, and~$\eta_1$ will depend on the plasma density and temperature, 
and so will vary from one point to another inside the reconnection layer.
In our simple reconnection model  we shall, however, ignore this aspect 
and assume that these parameters are constant in space and time and hence
the same profile $\eta(j)$ applies everywhere. In other words, 
$\eta$ is determined solely by the local value of~$j$.


\subsection{Two possible solutions and their stability}
\label{subsec-stability}

We shall now use the $\eta(j)$ dependence shown in Figure~\ref
{fig-resistivity} to find the correct value of~$j_0$ that corresponds
to the equilibrium Petschek solution. First, the requirement that 
$\Delta=Kl_\eta$ can be interpreted as follows. At $y=0$, let us have 
some value~$j_0$ and the corresponding value~$\eta(j_0)$. 
The requirement that $j_* \equiv j(l_\eta)=\xi j_0/e$ means 
that $\eta(y=l_\eta)=\eta(\xi j_0/e)$. But, by definition of 
$l_\eta$, $\eta(y=l_\eta)=\eta(j_0)/e$. Thus, $j_0$ is determined 
using function $\eta(j)$ from the condition:
\begin{equation}
\eta(j_0) = e\eta(j_*) = e \eta(\xi j_0/e) \, .
\label{eq-equilibrium}
\end{equation}

\begin{figure}
\plotone{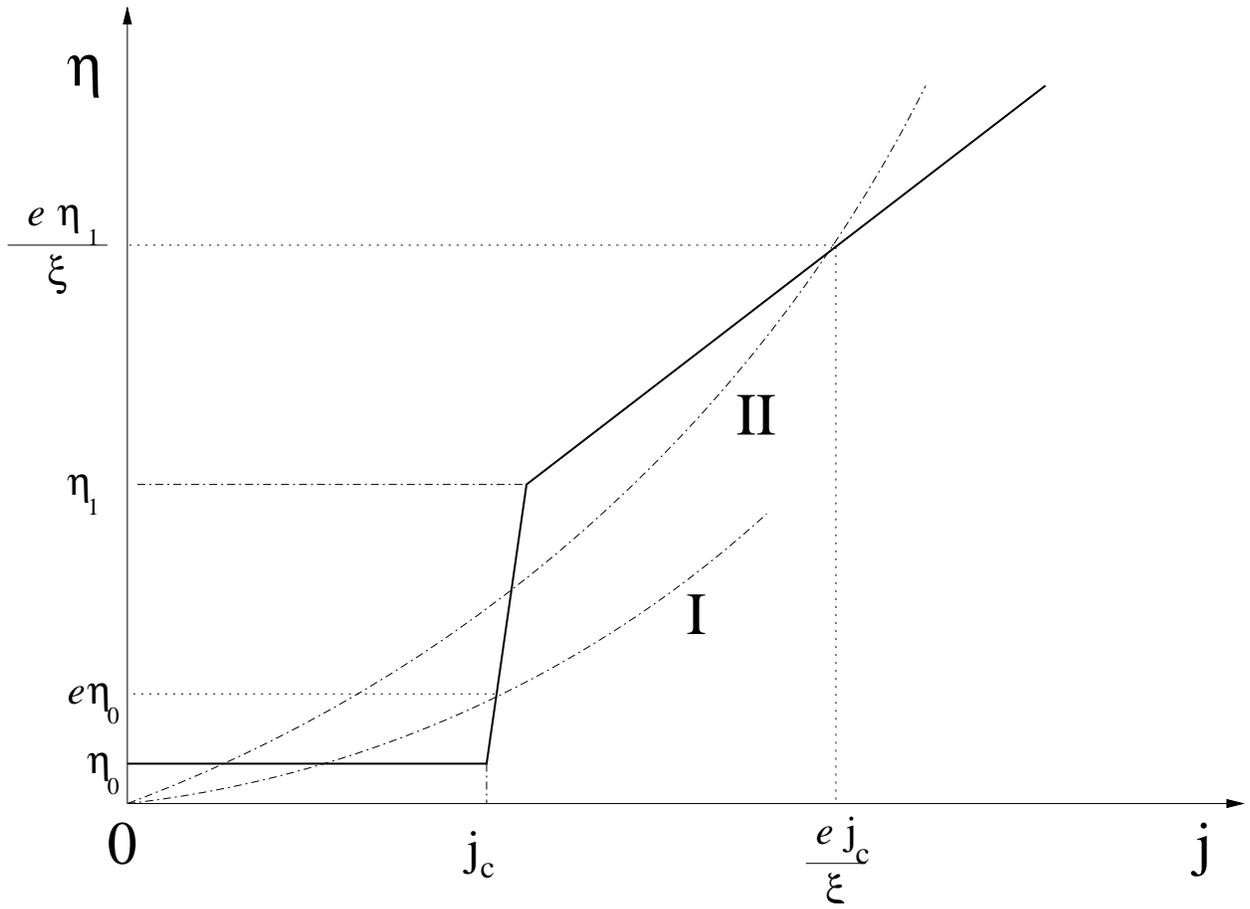}
\figcaption{The two possible equilibrium solutions.
\label{fig-solutions}}
\end{figure}

The solution of this equation for a given $\xi$ 
can be found by drawing a set of parabolae $\eta\propto j^\alpha$ 
where $\alpha = 1/(1-\ln\xi)$, and  selecting out of this set
those parabolae for which the points of their intersection with 
the curve~$\eta(j)$ are separated by a factor of~$e$ in their 
values of $\eta$, as shown in Figure~\ref{fig-solutions}.
It is clearly seen from this Figure that, for any $\xi>1$
(and, hence, $\alpha>1$) and for the general shape 
of~$\eta(j)$ adopted in this paper, one finds, in principle, 
two such solutions.

The first solution (curve~I) corresponds to~$j_0$ lying on 
the rapidly rising part of the $\eta(j)$ curve, $j_0^I \in 
[j_c,j_c+\Delta j]$, while $j_*^I=j^I(l_\eta)=\xi j_0/e <j_c$.
Then, $\eta^I(y=l_\eta)=\eta_0$ and hence $\eta^I(j_0)=e\eta_0$. 
We thus see that the corresponding resistivity enhancement in 
the inner diffusion region is not very large in this case, just 
by a factor~$e$ over the collisional value~$\eta_0$.

The second solution (curve~II) boasts much higher resistivity
enhancement. In this solution, $j_*^{II}$ falls within
the narrow rapidly rising part of the $\eta(j)$ curve, and
hence 
\begin{equation}
j_0^{II} = \xi^{-1} e j_*^{II} \simeq ej_c/\xi \, .
\label{eq-j0-II}
\end{equation}

Correspondingly, the resistivity at the center of the diffusion region is
\begin{equation}
\eta^{II}(j_0) =  \eta(j_0^{II}) =\eta_1 e/\xi \gg \eta_0 \, .
\label{eq-eta-II}
\end{equation}

Any Petschek-like configuration with~$j_0$ between~$j_0^I$ 
and~$j_0^{II}$, and hence with $e\eta_0<\eta<e\eta_1/\xi$,
will not be in a steady equilibrium and instead will evolve 
so as to increase~$j_0$. Indeed, if $\eta(j_0)>e\eta_0$, then 
$\eta(j_*) > \eta_0$ and hence~$j_* > j_c$. But then, since 
$j_0<ej_c/\xi$, we have $j_*/j_0>e/\xi$ and therefore $\Delta 
> K l_\eta$. As we know from numerical simulations, this leads to 
shrinkage of~$\Delta$. As we demonstrate below, this shrinkage, 
in turn, results in a decrease in~$\delta$ and hence in an 
increase in~$j_0$. The evolution will then presumably reach 
a stationary state when (the stable) solution~II is reached.

Now, which one of the two solutions I and~II will be realized in a real
physical system? We suggest that, in general, the answer to this question 
is determined by the stability of the solutions with respect to a small 
change in~$j_0$. In particular, we shall demonstrate that the first 
solution is unstable while the second one is stable. This will enable 
us to conclude that the system will evolve towards the second solution 
corresponding to higher $\eta_0$ and hence to a higher reconnection rate.

We shall first demonstrate that the stability properties are largely 
determined by the relative size of the slopes of the function $\eta(j)$ 
at $j = j_0$ and $j = j_* = \xi j_0/e$. We start by presenting a rather 
general stability analysis that works for any monotonically increasing 
function~$\eta(j)$. Hence, in order to apply this analysis to our 
solution~I, we need to modify our $\eta(j)$ slightly by assuming 
that this function has a non-zero positive (although arbitrarily 
small) slope $\eta'(j)>0$ in the region $j<j_c$. We then discuss 
a somewhat modified treatment for the case when $\eta'(j)$ is 
exactly zero in this region. These modifications are not essential 
and the conclusion regarding solution~I being unstable is the same.

So, first let us assume that $\eta'(j)\neq 0$ but has a small 
positive value at $j<j_c$. Let us consider an equilibrium 
Petschek-like configuration with $\Delta = K l_\eta$. Let this 
configuration be characterized by unperturbed values $\delta$, 
$j_0$, $j_*$, $\Delta$, $l_\eta$, etc. Now, imagine that at $t=0$ 
this equilibrium Petschek configuration is suddenly perturbed 
while preserving its Petschek-like character, that is changed 
into a neighboring Petschek configuration. This new Petschek 
configuration is, generally speaking, not in equilibrium, i.e.,
it does not satisfy condition~(\ref{eq-equilibrium}). It will 
then evolve through a sequence of Petschek states. For simplicity, 
here we envision the process as occurring in two parts with different 
timescales: the adjustment between~$\Delta$ and~$\delta$ to conform to 
the Sweet--Parker structure of the diffusion region is instantaneous,
whereas the adjustment of $\Delta$ to $l_\eta$ occurs on a longer timescale.
In reality this might not be true and these two processes may occur on the
same time scale (namely, the Alfv{\'e}n crossing time). But here we are 
only interested in the direction of the evolution, i.e., whether the 
perturbed system will move away or towards the equilibrium, so our 
qualitative analysis should be sufficient.

In particular, let us imagine that at $t=0$ the thickness~$\delta$ 
of the diffusion region is reduced slightly and hence the 
central current density is correspondingly slightly increased:
\begin{eqnarray} 
\delta &\rightarrow & \tilde{\delta} = \delta (1-\epsilon) \, , 
\label{eq-perturbation-delta} \\
j_0  &\rightarrow & \tilde{j_0} =  j_0 (1+\epsilon) \, , 
\label{eq-perturbation-j0} 
\end{eqnarray}
where $\epsilon \ll 1$.

An increase in $j_0$ by a factor $(1+\epsilon)$ leads to an
increase in the resistivity at the center:
\begin{equation}
\eta(j_0) \rightarrow \eta(\tilde{j_0}) =
\eta(j_0) + \eta'(j_0) j_0 \epsilon \, .
\label{eq-perturbation-eta}
\end{equation}

For the system to remain a valid Petschek solution,
we require that the inner diffusion region remains a Sweet--Parker
layer at all times, so the width $\Delta$ of the current distribution 
will automatically adjust to a new value, $\tilde{\Delta}(0)$, which, 
according to equation~(\ref{eq-SP-delta}), is
\begin{equation}
\Delta \rightarrow \tilde{\Delta}(0) = 
V_A {\tilde{\delta}^2\over{\eta(\tilde{j_0})}} =
\Delta\biggl[1-\epsilon\bigl(2+{{\eta'(j_0)j_0}\over{\eta(j_0)}}\bigr)\biggr]
\label{eq-perturbation-Delta}
\end{equation}

This equation expresses the initial reduction in~$\Delta$ 
in direct response to the reduction in~$\delta$ and hence 
to the related increase of~$j_0$ and $\eta(j_0)$ in the Sweet--Parker 
model for the diffusion region.
For example, if we are considering solution~I, $j_0\in[j_c,j_c+\Delta j]$,
then $\eta'(j_0)\simeq \eta_1/\Delta j \gg \eta_1/j_0 > \eta(j_0)/j_0$,
and hence the last term is dominant, i.e., the initial change in $\Delta$
is mainly due to the change in resistivity:
\begin{equation}
\tilde{\Delta}^I(0) \simeq \Delta \biggl( 1 -
\epsilon {{\eta'(j_0)j_0}\over{\eta(j_0)}} \biggr) \simeq \Delta
\biggl(1-{\eta_1\over{\eta(j_0)}} {j_c\over{\Delta j}} \epsilon \biggr)
\label{eq-perturbation-Delta-I}
\end{equation}

Similarly, if we are considering solution~II, then $j_0>j_c+\Delta j$,
and $\eta\propto j$ so that $\eta'(j_0)=\eta(j_0) j_0$. Then
\begin{equation}
\tilde{\Delta}^{II}(0) \simeq \Delta \biggl( 1- 3\epsilon \biggr) \, .
\label{eq-perturbation-Delta-II}
\end{equation}

Let us now ask what the new value $\tilde{l}_\eta$ of the width  
of the enhanced-resistivity region is and how it compares with 
the current sheet width~$\tilde{\Delta}(0)$.

From equation~(\ref{eq-perturbation-eta}) we see that the point 
$y=l_\eta$ moves to a location where $\eta(y=l_\eta)$ is increased by 
$\eta'(j_0) j_0\epsilon/e$ over its unperturbed value. On the 
other hand, this increment is equal to $\eta'(j_*) (\tilde{j_*}-j_*)$.
Therefore, $j_*$ has to increase by the amount 
\begin{equation}
\tilde{j_*} - j_* = {j_0\over e} \, \epsilon \, {\eta'(j_0)\over{\eta'(j_*)}}
= {\epsilon\over\xi} \, {\eta'(j_0)\over{\eta'(j_*)}} \, j_* \, .
\end{equation}

We have, by combining the expressions for the changes in~$j_0$ 
and~$j_*$,
\begin{equation}
{\tilde{j_*}\over\tilde{j_0}} = \biggl({j_*\over j_0} \biggr) \, 
\biggl[ 1+\epsilon \, \biggl( 
{1\over\xi}\, {{\eta'(j_0)}\over{\eta'(j_*)}} -1 \biggr)\biggr] \, .
\label{eq-perturbation-j*}
\end{equation}
We are now going to use this result to evaluate the change in~$l_\eta$.

Because of the symmetry with respect to $y=0$, the ratio $j(y)/j_0$ 
should be an even function of~$y$ with the characteristic scale~$\Delta$:
\begin{equation}
{{j(y)}\over j_0} = F(Y) \, ,
\label{eq-F(Y)}
\end{equation}
where 
\begin{equation}
Y\equiv {y\over\Delta} \, .
\label{eq-def-Y}
\end{equation}

Let us assume for simplicity that during the initial perturbation 
the shape of this function does not change. We can then write 
(here $Y_*\equiv l_\eta/\Delta$)
\begin{equation}
{\tilde{j_*}\over\tilde{j_0}} \equiv F(\tilde{Y_*}) = 
F(Y_*) \, \biggl[1 - \mu (\tilde{Y_*}-Y_*) \biggr]\, ,
\end{equation}
where 
\begin{equation} 
\mu \equiv - \biggl. {{F'(Y)}\over{F(Y)}} \biggr|_{Y=Y_*} = \, 
-  {e\over\xi} \, F'(Y_*) > 0 \, .
\end{equation}

By comparing this with the above result~(\ref{eq-perturbation-j*}) 
we see that the change in $Y_*$ is equal to
\begin{equation}
\tilde{Y_*}-Y_* = - {\epsilon\over\mu} \
\biggl[ {{\eta'(j_0)}\over{\xi\eta'(j_*)}} - 1 \biggr] \, .
\label{eq-perturbation-Y*}
\end{equation}

This equation tells us whether the ratio $Y_*=l_\eta/\Delta$ increases 
or decreases from its equilibrium value~$1/K$. We see that the result 
depends on the relative sizes of the slopes of function $\eta(j)$ at 
$j=j_0$ and $j=j_*$. In particular, in the case of solution~II,
$\eta'(j_0) < \xi \eta'(j_*)$, and then it follows from equation~(\ref
{eq-perturbation-Y*}) that~$l_\eta$ is reduced by a lesser factor 
than~$\Delta$. As a consequence, the resulting perturbed Petschek 
configuration has $\tilde{\Delta}<\tilde{\Delta}_{\rm eq}=K\tilde{l}_\eta$. 
We then expect, based on the results of numerical simulation mentioned 
above, that in this case $\tilde{\Delta}$ will increase 
towards~$\tilde{\Delta}_{\rm eq}$, thus negating the effect 
of the initial-perturbation decrease in~$\Delta$. 

In contrast, if $\eta'(j_0)>\xi\eta'(j_*)>0$, then~$l_\eta$ is reduced 
by a larger factor than~$\Delta$; thus the resulting perturbed Petschek 
configuration has $\tilde{\Delta}>\tilde{\Delta}_{\rm eq}=K\tilde{l}_\eta$. 
Invoking again the results of numerical simulations, we conclude that
the system will evolve in such a way as to decrease~$\tilde{\Delta}$ 
further. In fact, this scenario describes what happens in the case of
solution~I, but its application to this solution involves some subtlety.
Indeed, notice that one can use equations~(\ref{eq-perturbation-j*}) 
--- (\ref{eq-perturbation-Y*}) only if $\eta'(j_*)\neq 0$. Thus, in 
order to apply this general analysis to solution~I, we need to modify
the function $\eta(j)$ slightly to make it monotonically increasing in 
the region $j<j_c$. In other words we tentatively assume that this
function has a non-zero positive (although arbitrarily small) slope
$\eta'(j)>0$ in this region. This modification is not essential;
it needs to be introduced here purely for our technical convenience,
to be able to use our general equations~(\ref{eq-perturbation-j*}) 
--- (\ref{eq-perturbation-Y*}). We do not actually need to rely on it 
in order to derive the same conclusions regarding solution~I. Indeed, 
if we insist on our unmodified function $\eta(j)$ with the slope $\eta'(j)$ 
being exactly equal to zero for $j<j_c$, then, instead of 
equations~(\ref{eq-perturbation-j*})---(\ref{eq-perturbation-Y*}),
we can use an alternative argument to describe the response of solution~I
to the initial perturbation. This argument  is in fact even simpler and
physically more transparent than the one described above; it goes as follows.

Since $j_0$ in solution~I lies on the steeply rising part of the 
$\eta(j)$ dependence, the central resistivity is increased by a 
relatively large amount given by equation~(\ref{eq-perturbation-eta}).
On this part of the curve, we have $\eta'(j_0)\simeq \eta_1/\Delta j$, 
and so 
\begin{equation}
\eta(\tilde{j}_0) = e \eta_0 + \epsilon \eta_1 {j_0\over{\Delta j}}\, .
\label{eq-perturbation-eta-I}
\end{equation}
Then, $\eta(\tilde{l}_\eta)\equiv \eta(\tilde{j}_0)/e=\eta_0+ 
{\epsilon\over e}\eta_1 (j_c/\Delta j)$, and, therefore,
$j_*\in[j_c,\tilde{j}_0]$, i.e., $\tilde{j}_*\approx\tilde{j}_0$.
If the perturbed current density $\tilde{j}(y)$ has a characteristic
scale~$\tilde{\Delta}$, then $\tilde{j}_*\approx\tilde{j}_0$ implies 
that $\tilde{l}_\eta \ll\tilde{\Delta}$. Thus we see that, as the 
principal effect of the initial perturbation, the size~$\tilde{l}_\eta$ 
of the resistivity enhancement region drops very sharply, whereas all 
other quantities change rather smoothly. Then we can again invoke our 
empirical fact that a configuration like this will not stay stationary 
but will evolve so that the current layer width~$\tilde{\Delta}$ will 
tend to decrease to become comparable with~$\tilde{l}_\eta$, in agreement
with our expectations.

But as $\tilde{\Delta}$ changes during the subsequent evolution for $t>0$, 
there will be a feedback on~$\tilde{\delta}$. Indeed, since the diffusion 
region always remains a Sweet--Parker layer, $\tilde{\delta}$ 
and~$\tilde{\Delta}$ are always related via equation~(\ref{eq-SP-delta}). 
Since $\eta(j_0)$ is a monotonically increasing function of $j_0\propto 
1/\delta$, we see that $\tilde{\delta}$ is a monotonically increasing 
function of~$\tilde{\Delta}$. For example, for solution~I, we can, 
using the fact that $\tilde{j}_0$ is always very close to~$j_c$, 
approximate $\tilde{\delta}$ by~$\delta_c$ everywhere except where 
it appears in the combination $(\delta_c-\tilde{\delta})$. We thus 
can express the resistivity as $\eta(\tilde{j}_0) \simeq \eta_1 
(j_c/\Delta j) (\delta_c-\tilde{\delta})/ \delta_c$ and, substituting 
this expression into equation~(\ref{eq-SP-delta}), obtain
\begin{equation} 
\biggl({{\delta_c-\tilde{\delta}}\over\delta_c}\biggr)^I = 
{\delta_c\over{\tilde{\Delta}}} {{V_A\delta_c}\over\eta_1}
{{\Delta j}\over j_c} \, .
\label{eq-stability-I}
\end{equation}
We thus see that the decrease in~$\tilde{\Delta}(t)$ leads to a 
proportional increase in~$\delta_c-\tilde{\delta}(t)$. In a similar 
manner, for solution~II we find, using equations~(\ref{eq-SP-delta}) 
and~(\ref{eq-def-eta1}), that $\tilde{\delta}\propto\tilde{\Delta}^{1/3}$. 
In either case we see that after the initial perturbation, the subsequent 
evolution of the system is such that~$\delta$ changes in the same direction 
as~$\Delta$.

Now we are ready to apply the above general results to
the issue of stability of our solutions~I and~II.
We see that in the case of solution~I, the initial decrease
in~$\delta$ and~$\Delta$ tends to amplify further. This means
that this solution is unstable: a slight increase in~$j_0$ 
will lead to further increase and thus the system will move
away from the initial equilibrium, towards higher values of
central current density and resistivity, until it approaches 
solution~II.%
\footnote
{At the same time, a slight initial decrease in~$j_0$ will
presumably lead to a further decrease in~$j_0$ and hence an
increase in $\Delta$, until the system reaches the stable
Sweet--Parker configuration with $\Delta=L$ and $\eta(j_0)=\eta_0$.
This scenario, however, can only occur if the system's size~$L$ is
tremendously large; it is therefore of no importance to solar
flares, as discussed below [see equation~(\ref{eq-eta_min})].}
Solution~II, on the other hand, is stable; indeed, as have shown above, 
after its initial decrease [equation~(\ref{eq-perturbation-Delta-II})], 
$\Delta$ and hence~$\delta$ will tend to increase, thereby reducing the 
initial perturbation. Thus, in this case the feedback is negative and 
hence solution~II is stable.

It is important to realize that, in practice, solution~I is likely 
to be irrelevant for another reason (in addition to being unstable). 
Indeed, real astrophysical systems, including solar flares, always 
have a finite (albeit very large) global size~$L$. For the Petschek 
model to work, we must have $\Delta\leq L$. This condition imposes 
an upper limit on~$\Delta$ and hence, via equation~(\ref{eq-SP-delta}), 
a lower limit on~$\eta(j_0)$. For example, for~$j_0$ on the rapidly 
rising part of the~$\eta(j)$ curve, this condition can be cast as
\begin{equation}
\eta(j_0) > \eta_{\rm min}(L) \equiv {{\delta_c^2 V_A}\over L} \, .
\label{eq-eta_min}
\end{equation}
If $\eta_{\rm min}>e\eta_0$, then solution~I does not even exist
for a given global size~$L$. For typical solar corona conditions, 
collisional resistivity~$\eta_0$ is so small that inequality~(\ref
{eq-eta_min}) is not satisfied; in other words, one gets a very 
large value for $\Delta^I$, much larger than $L\simeq 10^9$~cm. 

Here is how the requirement that $\eta(j_0)>\eta_{\rm min}(L)$ comes 
into play from the evolutionary point of view. As a reconnection layer 
starts to form, its thickness decreases rapidly until it reaches~$\delta_c$; 
after that, $\delta$ and the central current density~$j_0$ both stay 
approximately constant while $\eta(j_0)$ increases rapidly. When 
$\eta(j_0)$ reaches $\eta_{\rm min}$, the corresponding width~$\Delta$ 
of the diffusion region, determined by equation~(\ref{eq-SP-delta}),
becomes equal to~$L$. From this moment on, the reconnection system can 
be described by the Petschek model. The particular Petschek configuration 
found exactly at this moment (i.e., when $\eta=\eta_{\rm min}$ and hence 
$\Delta=L$) is that of Sweet--Parker. If $\eta_{\rm min}>e\eta_0$, 
then $j(l_\eta)>j_c$ and hence is very close to~$j_0$; this means 
that resistivity is very strongly localized, $l_\eta\ll\Delta$. 
Therefore, this configuration will not be at equilibrium, and the 
system will evolve through a sequence of Petschek configurations 
with ever-increasing~$j_0$ and~$\eta(j_0)$. Each of these configurations 
with $j_0$ on the rapidly rising part of the $\eta(j)$ curve will have 
$\Delta> \Delta_{\rm eq}=Kl_\eta$ until the configuration corresponding 
to the stable equilibrium solution~II is reached.

We have thus demonstrated that a physical system with anomalous 
resistivity of the type shown on Figure 1 will evolve towards a stable 
solution II, which is characterized by the following values of the central 
current density and resistivity:
\begin{equation}
j_0 \simeq e j_c/\xi \, ,
\label{eq-j0}
\end{equation}
\begin{equation}
\eta(j_0) \simeq e \eta_1/\xi \, .
\label{eq-eta(j0)}
\end{equation}
The exact coefficients in these expressions require a precise 
model for $\eta(j)$, etc...

Finally, let us make a couple of remarks regarding our choice of~$\xi$. 
In the above considerations we have assumed that~$\xi>1$. Notice that 
the case $\xi=1$ ($\Delta_{\rm eq}=l_\eta$) is, in a certain sense, 
degenerate: in addition to an unstable solution~I, one gets a continuum 
of neutrally stable solutions~II that correspond to $j_0\geq ej_c$. Our 
present analysis does not allow us to discriminate among these solutions; 
they all appear to be equally plausible. If, however, one considers the 
unique stable solutions~II with $\xi>1$ and takes the limit $\xi
\rightarrow 1$, then one arrives at a single limiting solution 
(having $j_0=ej_c$). Thus, we use this limiting solution to extend 
our family of solutions with $\xi>1$ to include the $\xi=1$ case.

However, if $\xi<1$ ($\Delta_{\rm eq}<l_\eta$), then our present analysis
does not work: one still has an unstable solution I, but no solutions 
of type II exist, at least as long as $\eta(j)$ remains a linear function. 
This suggest that the system will continuously evolve towards higher and 
higher values of~$j_0$ and~$\eta(j_0)$ and hence the reconnection process
will continuously accelerate until some new resistivity saturation 
mechanism sets in. In this case, the system will stabilize and a 
steady state will be achieved at a much higher level of~$j_0$ than 
the level $j_0\sim j_c$ discussed here.


\subsection{Reconnection Rate}
\label{subsec-rate}

Thus, we now have a stable steady state Petschek-like configuration
with the central diffusion region characterized by the thickness
$\delta\simeq \delta_c/\zeta$ and the central resistivity $\eta=
\zeta\eta_1$. Here, $\zeta$ is a finite constant; according to 
our analysis, $\zeta=e/\xi$. However, we realize that the simple 
qualitative model presented here is not adequate for providing 
any numerical accuracy; it should only be used for order-of-magnitude 
estimates. Therefore, in the following we shall just keep using the 
above expressions for~$\delta$ and~$\eta(j_0)$, parameterizing our 
ignorance by a constant $\zeta=O(1)$.

Let us now ask, what is the reconnection rate associated with
this configuration? Using expression~(\ref{eq-SP-scaling}) we 
find the aspect ratio of the central diffusion region:
\begin{equation}
{\Delta\over\delta}= {V_A\delta\over{\eta(j_0)}}= {S_*\over{\zeta^2}} \, ,
\label{eq-aspect-ratio}
\end{equation}
where we define 
\begin{equation}
S_* \equiv {{V_A\delta_c}\over\eta_1}  \, .
\label{eq-def-S*}
\end{equation}
[Note: the parameter $S_*$ defined here is completely different 
from Kulsrud's (2001)~$S^*$.]

The reconnection velocity is (assuming that the density stays
roughly constant inside the reconnection layer):
\begin{equation}
{V_{\rm rec}\over V_A} = {\delta\over\Delta} = \zeta^2 S_*^{-1} \, ,
\label{eq-V_rec}
\end{equation}
and the typical reconnection timescale is 
\begin{equation}
\tau_{\rm rec} \equiv {L\over{V_{\rm rec}}} = \tau_A(L)\, S_*/\zeta^2 \, ,
\label{eq-def-tau_rec}
\end{equation}
where $\tau_A(L)\equiv L/V_A$ is the global Alfv{\'e}n crossing time.

Note that the aspect ratio~(\ref{eq-aspect-ratio}) and hence 
the reconnection velocity~(\ref{eq-V_rec}) turn out to be 
independent of the global system size~$L$, but only depend 
on the local plasma parameters.

Also note that our expression~(\ref{eq-V_rec}) for the reconnection 
velocity differs from the result $V_{\rm rec}/V_A \sim (\delta_c
\eta^*/V_A L^2)^{1/3}$ obtained previously by Kulsrud [see equation~(26) 
of Kulsrud 2001]. We believe that this discrepancy can be attributed, 
at least partly, to a somewhat different functional form of the 
$\eta(j)$-dependence adopted in his paper.


\section{Petschek Reconnection in the Presence of Anomalous
Resistivity due to Ion-Acoustic Turbulence (IAT)}
\label{sec-IAT}

In this section we assume that the anomalous resistivity enhancement 
is caused by the scattering of the current-carrying electrons off 
ion-acoustic waves, which are themselves excited by the ion-acoustic  
instability when the current density exceeds a certain threshold. 
The theory of ion-acoustic turbulence and the associated with it 
anomalous resistivity has greatly progressed over the past 40 years 
(e.g., Kadomtsev 1965; Rudakov \& Korablev 1966; Sagdeev 1967; 
Tsytovich \& Kaplan 1971; Biskamp \& Chodura 1972; Coroniti \& 
Eviatar 1977; Bychenkov~et~al. 1988). This theory is now very mature
and seems to be capable of producing reliable quantitative results 
regarding anomalous resistivity. In this paper,
we use the results presented by Bychenkov~et~al. (1988).%
\footnote
{We note, however, that here we use the results of the theory of 
ion-acoustic anomalous resistivity that has been developed for a 
homogeneous plasma without magnetic field. We acknowledge that 
the resistivity may be modified by both the presence of the 
magnetic field and by the fact that in our analysis turbulence 
is presumed to be confined to the very small central diffusion 
region, and is thus strongly inhomogeneous.}

Following the analysis of ion-acoustic turbulence by
Bychenkov~et~al. (1988), we adopt
\begin{equation}
j_c = a_1 en_e v_s \, ,
\label{eq-j_c}
\end{equation}
where $v_s \equiv \sqrt{Z T_e/m_i}$ is the ion sound speed,
and $a_1=O(1)$. In particular, according to equation~(2.143) 
of Bychenkov et al. (1988), $a_1 \simeq 2.14$.

Further, we compute anomalous resistivity $\eta(j>j_c+\Delta j)$ 
by using the expression~(2.148) of Bychenkov~et~al. (1988) 
for~$\sigma=\sigma_{\rm anom}(E)$ to express $\eta=c^2/4\pi\sigma$ 
in terms of the electric current density~$j=\sigma E$. We thus obtain%
\footnote
{Note that the value for anomalous conductivity given by 
Bychenkov~et~al. (1988) differs by a numerical factor of
order one from the famous Sagdeev's (1967) formula.}
\begin{equation}
\eta(j) \simeq {1\over{4\pi\cdot 0.16}}\, {c^2\over{\omega_{pe}^2}}\, 
\biggl({\lambda_{De}\over{\lambda_{Di}}}\biggr)^2 \, 
{j\over\sqrt{8\pi n_e T_e}} \simeq
{1\over 36}\, {c^2\over\omega_{pe}}\, {ZT_e\over T_i}\, 
\sqrt{Zm_e\over m_i}\, {j\over{en_e v_s}} \, .
\label{eq-eta-anom}
\end{equation}
We then estimate $\eta_1$ by extrapolating this dependence down 
to $j=j_c+\Delta j \simeq j_c$. Using equation~(\ref{eq-j_c}),
we get
\begin{equation}
\eta_1=\eta(j_c) = a_2 \, {c^2\over\omega_{pe}}\, {ZT_e\over T_i}\, 
\sqrt{Zm_e\over m_i}\, ,
\label{eq-eta1}
\end{equation}
where 
\begin{equation}
a_2 \simeq {a_1\over 36} \simeq 0.06 \, .
\label{eq-a2}
\end{equation}

For a pure hydrogen plasma ($Z=1$, $m_i=m_p$), we have:
\begin{equation}
\eta_1 = 1.4\cdot 10^{-3} {c^2\over\omega_{pe}}\, {T_e\over T_i}\, .
\label{eq-eta_1-hydrogen}
\end{equation}

As for $\eta_0$, we take it to be the classical collisional
Spitzer resistivity 
\begin{equation}
\eta_0 \equiv \eta_{\rm Sp} = {c^2\over{4\pi\sigma_{\rm Sp}}} \simeq
{{0.02 \Lambda}\over N_D} {c^2\over\omega_{pe}} \, ,
\label{eq-eta-Spitzer}
\end{equation}
where $\Lambda$ is the Coulomb logarithm and 
\begin{equation}
N_D\equiv n_e \lambda_{De}^3 \, ,
\label{eq-def-N_D}
\end{equation}
$\lambda_{De}\equiv\sqrt {T_e/4\pi n_e e^2}$ being the electron 
Debye radius. We see that we may expect a potential resistivity 
enhancement on the order of
\begin{equation}
{\eta_1\over\eta_{Sp}} = {N_D\over{2\Lambda}} {T_e\over T_i}
\sqrt{m_e\over m_i} \gg 1.
\label{eq-resistivity-enhancement}
\end{equation}

For this model of anomalous resistivity, $\delta_c$ and $\eta_1$ 
are given by expressions~(\ref{eq-def-delta_c}) and~(\ref{eq-eta1}) 
and then we can express~$S_*$ as
\begin{equation}
S_* = {1\over{a_1 a_2}}\, {V_A^2\over{cv_s}}\, {T_i\over ZT_e}\, 
{m_i\over Zm_e} \, .
\label{eq-S*-1}
\end{equation}

Expressing the ratio~$V_A/v_s$ in terms of the composite 
electron plasma beta parameter,
\begin{equation}
\beta_e \equiv {{8\pi n_e T_e}\over{B_0^2}}\, ,
\label{eq-def-beta_e}
\end{equation}
as $V_A/v_s=\sqrt{2/\beta_e}$, we can rewrite~(\ref{eq-S*-1}) as
\begin{equation}
S_* = a_3 {V_A\over{c\sqrt{\beta_e}}}\, 
{T_i\over ZT_e}\, {m_i\over Zm_e}\, ,
\label{eq-S*-2}
\end{equation}
where 
\begin{equation}
a_3 \equiv {\sqrt{2}\over{a_1 a_2}} \simeq 11 \, ,
\label{eq-def-a3}
\end{equation}
where we substituted $a_1=2.14$ and made use of equation~(\ref{eq-a2}).

Note that from equations~(\ref{eq-def-delta_c}) and~(\ref{eq-j_c})
it easily follows that 
\begin{equation}
{\delta_c\over{d_i}} = {1\over a_1} \sqrt{2\over\beta_e}\, ,
\label{eq-delta_c-d_i}
\end{equation}
where 
\begin{equation}
d_i \equiv {c\over{\omega_{pi}}}
\label{eq-def-d_i}
\end{equation}
is the ion skin depth. Substituting this very useful expression, along 
with equation~(\ref{eq-S*-2}), into our equation~(\ref{eq-aspect-ratio}) 
for~$\Delta$, we find
\begin{equation}
\Delta \simeq {a_4\over{\zeta^3}}\, d_i \, {m_i\over{Zm_e}}\, 
{T_i\over{ZT_e}}\, {V_A\over{c\beta_e}} \, ,
\label{eq-Delta}
\end{equation}
where 
\begin{equation}
a_4\equiv {2\over{a_1^2 a_2}} \simeq 7.3 \, .
\label{eq-def-a4}
\end{equation}

We also get an expression for the Alfv{\'e}n crossing time
for the diffusion region --- one of the most important
timescales in the problem:
\begin{equation}
\tau_A(\Delta) \equiv {\Delta\over V_A} \simeq 
{a_4\over{\zeta^3}}\, \omega_{pi}^{-1}\, \beta_e^{-1} \, 
{m_i\over{Zm_e}}\, {T_i\over{ZT_e}} \, .
\label{eq-tau_A}
\end{equation}

In addition, substituting equation~(\ref{eq-S*-2}) into equation~(\ref
{eq-def-tau_rec}) for the reconnection time~$\tau_{\rm rec}$, we find 
a very simple relationship expressing $\tau_{\rm rec}$ in terms of the 
light crossing time~$L/c$ and~$\beta_e$:
\begin{equation}
\tau_{\rm rec} \simeq {a_3\over{\zeta^2 \sqrt{\beta_e}}} \, 
{L\over c} \, {T_i\over{ZT_e}}\, {m_i\over{Zm_e}}  \, .
\label{eq-tau_rec-2}
\end{equation}

For a pure hydrogen plasma ($Z=1$, $m_i=m_p=1836\, m_e$), we get
\begin{eqnarray}
S_* &\simeq & 2\cdot 10^4 {V_A\over{c\sqrt{\beta_e}}}\, {T_i\over T_e}\, , 
\label{eq-S*-hydrogen} \\
\Delta &\simeq& 1.3\cdot 10^4 \zeta^{-3} \, d_i {T_i\over T_e}\, 
{V_A\over{c\beta_e}}\, , 
\label{eq-Delta-hydrogen} \\
\tau_A(\Delta) &\simeq& 1.3\cdot 10^4 \zeta^{-3}\, \omega_{pi}^{-1} \, 
{T_i\over T_e}\, \beta_e^{-1} \, , 
\label{eq-tau_A-hydrogen} \\
\tau_{\rm rec} &\simeq & {{2\cdot 10^4}\over{\zeta^2 \sqrt{\beta_e}}} \, 
{L\over c} \, {T_i\over T_e} \, .
\label{eq-tau_rec-hydrogen}
\end{eqnarray}

Note that in all these expressions $n_e$ and $T_e$ are to be taken
at the {\it center} of the reconnection layer ($x=y=0$), while the 
magnetic field~$B_0$ is the reconnecting magnetic field {\it outside} 
the layer, at~$x>\delta$, $y=0$.


\section{Application to Solar Flares}
\label{sec-flares}

Let us now try to apply the above results to typical solar flare 
conditions and see whether our model is able to explain the very 
short time scale of impulsive flares. In order to be able to make 
quantitative estimates, we shall first need to discuss the values 
of some relevant plasma parameters.

Table~\ref{table} lists the values of the key parameters of our model,
along with the values of some fundamental plasma parameters, for two 
sets of conditions. Both sets are calculated for fully-ionized pure 
hydrogen plasma. The first set (column~III) illustrates the fiducial 
solar coronal conditions: $B_0=100$~G, $n_e=10^9$~cm$^{-3}$, and 
$T_e=T_i=2\cdot 10^6\, K \simeq 200$~eV. The second set (column~IV) 
corresponds to the fiducial solar flare conditions which will be 
discussed below.

For the parameters in column~III we see that the characteristic 
reconnection time turns out to be no more than an order of magnitude
longer than the observed flare duration time (which is typically of 
order $10^3$~sec), but it is, apparently, still not sufficiently short. 
Notice, however, that the fiducial solar corona parameters used in
column~III may not be appropriate for the center of the solar flare 
reconnection layer. Indeed, one can expect that the turbulence will 
lead to rapid heating of the plasma, resulting in a substantial rise 
of the electron temperature.

We can then ask whether the electron thermal pressure at the center 
of the current layer will grow to a level where it becomes comparable 
to the outside magnetic pressure, $\beta_e=O(1)$. This is a very 
important question because, according to equation~(\ref{eq-delta_c-d_i}), 
the value of~$\beta_e$ controls the regime that the system finds itself 
in. In particular, if $\delta<d_i$, then the diffusion region needs to 
be described in terms of electron MHD (or Hall MHD), the theory of which 
in the reconnection context has recently been greatly advanced by a number 
of researchers (e.g., Drake~et~al. 1994; Biskamp~1997; Bhattacharjee~et~al.
2001). Thus, we see that~$\beta_e$ is a very important parameter, whose 
value may have a profound influence on the applicability of the anomalous 
resistivity model adopted in this paper. Let us now ask what value 
for~$\beta_e$ one can expect in the case of magnetic reconnection 
in solar corona. 

First, notice that it seems inevitable that, if there is no axial 
(or guide) magnetic field component, that is when $B_z=0$, then one 
has to have $\beta_e=O(1)$. Indeed, in this case the basic requirement 
of pressure balance across the current layer dictates that the plasma 
pressure at the center of the layer's diffusion region be equal to the 
pressure of the magnetic field outside of the layer. This means that 
$\beta_e=1$ when $T_e\gg T_i$ (and hence the ion pressure inside the 
layer is negligible) and $\beta_e=1/2$ when $T_i=T_e$ (and hence the 
ion pressure is equal to the electron pressure for pure hydrogen plasma). 
This conclusion holds regardless of the energy budget balance. For 
example, if there is no effective cooling mechanism, then the electron 
density~$n_e$ does not change significantly (i.e., by more than a factor
of order one), whereas the temperature increases up to the ``equipartition 
level'', $T_{\rm eq}=B_0^2/8\pi n_e$, which is about $3\cdot 10^8$~K for 
$n_e=10^{10}$~cm$^{-3}$ and $B_0=100$~G. On the other hand, if there is 
some effective cooling, the plasma temperature cannot reach such a high 
value and then the density is increased instead to maintain the pressure 
balance across the layer. In either case, one finds $\beta_e=O(1)$ and 
hence, according to equation~(\ref{eq-delta_c-d_i}), $\delta \sim d_i$. 
This means that the system will require a Hall-MHD description as soon 
as, or even before, the anomalous resistivity becomes important.

In the case of solar flares, however, it is unclear whether 
this situation is present. Indeed, there exist a possibility
for maintaining the pressure balance across the layer with
$\beta_e\ll 1$. This scenario requires two things, both likely 
to be relevant in solar flare environment: the presence of a 
non-zero axial component of the magnetic field~$B_z$ and some 
plasma cooling mechanism. Indeed, when the plasma is cooled 
efficiently, the increase in thermal pressure that is required 
to maintain the pressure balance cannot come from the increase 
in temperature, and so the plasma tends to compress inside the 
reconnection layer. This, in turn, leads to the proportional 
compression of the guide field component (here we are neglecting
the resistive decoupling between the guide field and the plasma):
\begin{equation}
{{B_z\vert_{\rm inside}}\over{B_z\vert_{\rm outside}}} = 
{{n_e\vert_{\rm inside}}\over{n_e\vert_{\rm outside}}} \, .
\end{equation}
(Here the subscript ``inside''corresponds to the center 
of the reconnection layer, $x=y=0$, whereas the subscript 
``outside'' corresponds to the plasma above the layer,
$y=0$, $x \gg \delta$.)

If cooling is so strong that the resulting central temperature
is small compared with the equipartition temperature, then the 
plasma pressure can be neglected in the pressure balance. The 
pressure balance is then achieved with the increased guide field 
pressure inside the layer balancing the reconnecting field's 
pressure outside the layer:
\begin{equation}
B_z^2\vert_{\rm inside} - B_z^2\vert_{\rm outside} = B_0^2 \, .
\label{eq-pressure-balance}
\end{equation}

Thus, $B_z\vert_{\rm inside}$ is determined from the pressure 
balance~(\ref{eq-pressure-balance}) and then the ratio 
$B_z\vert_{\rm inside}/B_z\vert_{\rm outside}$ determines the 
compression factor and hence the central density $n_e\vert_{\rm 
inside}$. Typically one might expect $B_z\vert_{\rm outside} 
\sim B_0$, and so~$B_z$ (and hence~$n_e$) is increased at the 
center of the layer by a factor of order one. As for the central 
electron temperature, it is going to be determined by the balance 
between the turbulent ohmic heating and the cooling due to electron 
thermal conduction.%
\footnote
{In the context of solar flares the radiative cooling of 
the diffusion region, including both the bremsstrahlung 
and cyclotron mechanisms, appears to be ineffective, as 
the characteristic radiative cooling time is much longer 
than the time $\tau_A(\Delta)$ that a fluid element spends 
inside the diffusion region.}
This is an important and very complicated problem and its
detailed treatment lies outside the scope of this paper.
Therefore, here we shall give only some very simple estimates.

The characteristic ohmic heating time can be evaluated as
\begin{equation}
\tau_{\rm heat} \sim {{n_e T_e}\over Q}\, ,
\label{eq-tau_heat-1}
\end{equation}
where the ohmic heating rate per unit volume roughly is 
\begin{equation}
Q\sim {{j_0^2}\over\sigma} \sim 
{{B_0^2}\over{4\pi}} {{\eta(j_0)}\over\delta^2} \sim 
{{B_0^2}\over{4\pi}} {1\over{\tau_A(\Delta)}}\, ,
\label{eq-heating-rate}
\end{equation}
so that 
\begin{equation}
\tau_{\rm heat} \sim \tau_A(\Delta) \beta_e \, ,
\label{eq-tau_heat-2}
\end{equation}
where we have dropped numerical factors of order unity.
We thus see again that, in the case where there is no cooling, 
a cold fluid element entering the reconnection layer will be 
heated up very rapidly; in fact, the total time that the element 
spends (and is being heated) inside the inner diffusion region
[of order $\tau_A(\Delta)$] is long enough for the plasma thermal 
pressure to reach $\beta_e=O(1)$.

If there is efficient cooling due to electron thermal transport, 
then an absolute lower bound on the electron cooling time is set 
by the time it takes a freely streaming thermal electron to leave 
the inner diffusion region of size~$\Delta$:
\begin{equation}
\tau_{\rm cool}^{\rm min} \sim {\Delta\over{v_{\rm th, e}}} \sim
\tau_A(\Delta) {1\over\sqrt{\beta_e}} \sqrt{m_e\over m_i} \, .
\label{eq-tau_cool-min}
\end{equation}
Then, equating $\tau_{\rm cool}^{\rm min}$ and $\tau_{\rm heat}$, 
we get a lower bound on~$\beta_e$:
\begin{equation}
\beta_e^{\rm min} \sim \biggl({m_e\over m_i}\biggr)^{1/3} \ll 1 \, .
\label{eq-beta_e-min}
\end{equation}

This is the regime that is illustrated in column~IV of Table~\ref{table}.
Here, we use the following values of plasma parameters for our ``fiducial 
solar flare conditions'': 
$B_0=100$~G, $n_e=10^{10}$~cm$^{-3}$, $T_e=3\cdot 10^7\, K 
\simeq 3000$~eV, and $T_i=3\cdot 10^6\, K \simeq 300$~eV, 
which correspond to $\beta_e \simeq 0.1$. The resulting 
reconnection time scale $\tau_{\rm rec}$ is of the order
of a hundred seconds, which is fast enough to explain the 
observed very short duration of the impulsive phase of 
solar flares.

An upper limit on $\tau_{\rm cool}$ is given simply by the time 
that a fluid element spends inside the inner diffusion region, 
i.e., the Alfv{\'e}n transit time~$\tau_A(\Delta)$. This corresponds 
to the upper limit $\beta_e^{\rm max}=O(1)$, just as in a situation 
without a guide magnetic field or without cooling. The electron 
temperature then grows to about the equipartition value $T_{\rm eq}
\simeq 3\cdot 10^8$~K (for $n_e=10^{10}$~cm$^{-3}$ and~$B_0=100$~G).

Here are a few more numbers:\\
23, 45.32, 18650, -0.652.  :)

Thus, we can constrain $\beta_e$ to lie between $\beta_e^{\rm min}\sim 
10^{-1}$ and $\beta_e^{\rm max}=O(1)$. We then see from equation~(\ref
{eq-delta_c-d_i}) that even in the case of the lowest possible~$\beta_e$, 
the thickness~$\delta$ of the inner reconnection layer is roughly of the 
same order of magnitude as~$d_i$. This suggests that the Hall-MHD regime 
is likely to be at least marginally important in the physics of solar flares.
In this case the nature of anomalous dissipative processes may differ from 
the simple effective resistivity $\eta(j)$ due to ion-acoustic turbulence 
as described in \S~\ref{sec-IAT} and thus a more elaborate theory is needed.

Finally let us address one more question related to the applicability
of the ion-acoustic regime. One can raise the objection that, in order
to excite ion-acoustic instability, the condition $T_e\gg T_i$ needs to 
be satisfied; if instead the plasma is nearly isothermal, with $T_e\sim 
T_i$, then Buneman instability (Buneman 1959) can in principle be excited, 
but at a higher current-density threshold, $j_c^{\rm Buneman}\sim en_e 
v_{\rm th,e}\gg en_e v_s$. To address this problem, let us consider a 
plasma that initially (i.e., before the onset of reconnection) is 
isothermal. As the reconnection current layer is been formed, the 
current density in the layer gradually increases and finally reaches 
the Buneman instability threshold (calculated for the initial, relatively 
low temperature $T_{e,0}\sim 2\cdot 10^6$ K). The subsequent development 
of the instability leads to anomalous turbulent heating (Sagdeev~1967; 
Biskamp \& Chodura 1973; Bychenkov~et~al. 1988; Kingsep~1991), which 
raises the electron temperature faster than the ion temperature. 
At some point, the electron temperature becomes much higher than 
the ion temperature and the ion-acoustic instability is excited. 
Such a transition from the Buneman regime to the ion-acoustic regime 
has in fact been studied previously (see Bychenkov~et~al. 1988 and 
references therein). In magnetic reconnection context, a possibility 
of this transition has recently been discussed briefly by Roussev~et~al. 
(2002). One complication, however, is that, since the plasma is constantly 
moving through the vicinity of the neutral point, then, in order to sustain 
the IAT, one needs to pre-heat the electrons (relative to the ions) in every 
fluid element that is just entering the diffusion region. This means that 
some sort of anomalous heat leakage across the magnetic field is probably 
needed. Whether and how this can be achieved is a difficult question, which
falls outside of the scope of the present study.


\section{Conclusions}
\label{sec-conclusions}

In this paper we have presented a model of magnetic reconnection
in the presence of a current-driven enhanced anomalous resistivity.
This is a very simplistic, crude model that aims at predicting
the qualitative behavior of the system and the scaling of the 
reconnection rate with various plasma parameters, while treating 
numerical factors of order one only very approximately.

In this model we have combined the following three ingredients.
The first one is the observation, derived from several recent 
resistive-MHD numerical simulations (Erkaev~et~al. 2000, 2001; 
Biskamp \& Schwarz 2001), that whenever the resistivity is strongly 
localized, the reconnecting system will develop a Petschek-like 
configuration, with the width of the inner diffusion region of 
the order of the resistivity localization scale. The second 
ingredient of our model is the Sweet--Parker model (Sweet~1958;
Parker 1957, 1963) for the diffusion region of a Petschek 
configuration (Petschek~1964). Finally, the third ingredient 
is a physically realistic model for a current-driven anomalous 
resistivity expressed as a function~$\eta(j)$, which exhibits 
two characteristic features. The first feature is a sudden jump 
of~$\eta$ from a small collisional value~$\eta_0$ to a much larger 
value~$\eta_1$ as soon as~$j$ exceeds a threshold~$j_c$. The second
feature is a subsequent linear growth $\eta\propto j$ for~$j>j_c$.
This choice is motivated by the theory of anomalous resistivity 
due to ion-acoustic turbulence, which has been developed in detail 
over the last 40 years (see, e.g., Bychenkov et al. 1988).

Note that the anomalous resistivity function adopted in this paper 
becomes very sensitive to electric current density when the latter 
exceeds some threshold value~$j_c$; this makes it possible for the 
resistivity to be enhanced only in a small region, which, in turn, 
leads to the development of a Petschek-like configuration. Thus, 
our model is characterized by a reconnection rate that is enhanced 
(with respect to the classical, collisional-resistivity Sweet--Parker 
rate) by a combined action of anomalous resistivity and of the Petschek 
mechanism. It is important to realize that the role of anomalous 
resistivity in the acceleration of the reconnection process is two-fold: 
in addition to its direct action (lowering the global Lundquist number 
$S\equiv V_A L/\eta$), it accelerates reconnection indirectly, by 
turning on the Petschek mechanism.

The width of the inner diffusion region of the Petschek model,
and thus the resistivity localization scale, are determined 
self-consistently when all the ingredients of the model are 
taken into account.

Based on our stability analysis of two possible Petschek-like 
states, we predict that the system will evolve towards a certain 
stable Petschek-like configuration. This stable configuration is 
characterized by the central current density $j_0$ and the central 
resistivity $\eta(j_0)$ exceeding $j_c$ and $\eta_1$, respectively,
by a finite factor of order one. The reconnection velocity then 
scales as $V_{\rm rec} \sim V_A/S_*=\eta_1/\delta_c$, where 
$\delta_c=cB_0/4\pi j_c$ is the critical thickness of the layer.

We then consider (in \S~\ref{sec-IAT}) the case of anomalous 
resistivity due to ion-acoustic turbulence, as an important 
specific example. We derive very simple expressions for the 
parameters of the reconnection system (e.g., the width~$\Delta$ 
of the diffusion region, reconnection velocity $V_{\rm rec}$,
and the reconnection time scale $\tau_{\rm rec}$) in terms
of the basic plasma parameters $n_e$, $T_e$, $T_i$, and~$B_0$.

Finally, in \S~\ref{sec-flares}, we apply our model to the solar 
flare environment. We note that reconnection process will lead 
to significant electron heating, so that the electron pressure 
at the center of the reconnection layer may become comparable 
to the pressure of the reconnecting magnetic field outside the 
layer. Based on our model, we obtain typical reconnection times 
of order $10^2-10^3$ sec; this is short enough to explain the 
very fast time scale of impulsive flares. We note however that, 
as a result of the plasma heating inside the reconnection layer, 
the thickness $\delta$ of the diffusion region quickly becomes 
comparable to, or even smaller than, the ion skin-depth, $d_i
\equiv c/\omega_{pi}$. At these scales, new physical processes, 
described by Hall MHD, may come into play (Drake~et~al. 1994; 
Biskamp~1997; Bhattacharjee~et~al. 2001) even before the IAT 
develops and anomalous resistivity becomes important.

I am grateful to S.~Boldyrev, H.~Li, R.~Kulsrud, and R.~Rosner
for some very helpful discussions and interesting comments.
I would like to acknowledge the support by the NSF grant 
NSF-PHY99-07949.


\section*{REFERENCES}
\parindent 0 pt

Bhattacharjee, A., Ma, Z.~W., \& Wang, X. 2001, Phys. Plasmas,
8, 1829.

Biskamp, D. 1986, Phys. Fluids, 29, 1520.

Biskamp, D. 1997, Phys. Plasmas, 4, 1964.

Biskamp. D. 2000, ``Magnetic Reconnection in Plasmas'', 
(Cambridge Univ. Press, New York, 2000).

Biskamp, D. \& Chodura, R. 1973, Phys. Fluids, 16, 888.

Biskamp, D. \& Schwarz, E. 2001, Phys. Plasmas, 8, 4729. 

Buneman, U. 1959, Phys. Rev., 115, 503.

Bychenkov, V.~Yu., Silin, V.~P., \& Uryupin, S.~A. 1988,
Phys. Reports, 164, 121.

Coppi, B., \& Friedland, A.~B. 1971, ApJ, 169, 379.

Coroniti, F.~V. \& Eviatar, A. 1977, Ap. J.Suppl., 33, 189.


Drake, J.~F., Kleva, R.~G., \& Mandt, M..~E. 1994, 
Phys. Rev. Lett., 73, 1251.

Erkaev, N.~V., Semenov, V.~S., \& Jamitzky, F. 2000,
Phys. Rev. Lett., 84, 1455.

Erkaev, N.~V., Semenov, V.~S., Alexeev, I.~V., \& Biernat, H.~K. 2001,
Phys. Plasmas, 8, 4800.

Giovanelli, R.G. 1946, Nature, 158, 81.

Ji, H., Yamada, M., Hsu, S., \& Kulsrud, R. 1998, Phys. Rev. Lett.,
80, 3256.

Ji, H., Yamada, M., Hsu, S., Kulsrud, R., Carter, T., \& Zaharia, S. 1999,
Phys. Plasmas, 6, 1743.

Kadomtsev, B.~B. 1965, Plasma Turbulence (New York: Academic Press)

Kadomtsev, B.~B. 1975, Sov. J. Plasma Phys., 1, 389.

Kingsep, A.~S. 1991, Sov. J. Plasma Phys., 17, 342

Kulsrud, R.~M. 1998, Phys. Plasmas, 5, 1599.

Kulsrud, R.~M. 2001, Earth, Planets and Space, 53, 417.

Parker, E.~N. 1957, J. Geophys. Res., 62, 509.

Parker, E.~N. 1963, ApJ Supplement, 8, 177.

Petschek, H.~E. 1964, AAS-NASA Symposium on Solar Flares, 
(National Aeronautics and Space Administration, Washington, 
DC, 1964), NASA SP50, 425.

Rosner, R., Golub, L., Coppi, B., \& Vaiana, G.~S. 1978,
ApJ, 222, 317

Roussev, I., Galsgaard, K., \& Judge, P.~G. 2002, A\&A, 382, 639.

Rudakov, L.~E. \& Korablev, L.~V. 1966, Sov. Phys. --- JETP, 23, 145.

Sagdeev, R.~Z. 1967, Proc. Symp. Appl. Math., 18, 281.

Sato, T. \& Hayashi, T. 1979, Phys. Fluids, 22, 1189.

Scholer, M. 1989, J. Geophys. Res., 94, 8805.

Smith, P.~F. \& Priest, E.~R. 1972, ApJ, 176, 487.

Sweet, P.~A. 1958, in ``Electromagnetic Phenomena in Cosmical Physics'', 
ed. B.Lehnert, (Cambridge University Press, New York, 1958), p.~123.

Tsuneta, S. 1996, ApJ, 456, 840. 


Ugai, M. \& Tsuda, T. 1977, J. Plasma Phys., 17, 337.

Ugai, M. 1986, Phys. Fluids, 29, 3659.

Ugai, M. 1992, Phys. Fluids B, 4, 2953.

Ugai, M. 1999, Phys. Plasmas, 6, 1522.

Ugai, M. \& Kondoh, K. 2001, Phys. Plasmas, 8, 1545.


Uzdensky, D.~A. \& Kulsrud, R.~M. 2000, Phys. Plasmas, 7, 4018.

Vasyliunas, V.~M. 1975, Rev. Geophys. Space Phys., 13, 303.

Yamada, M., Levinton, F. M., Pomphrey, N., Budny, R., 
Manickam, J., \& Nagayama, Y. 1994, Phys. Plasmas, 1, 3269.

Yamada, M., Ji, H., Hsu, S., Carter, T., Kulsrud, R., Bretz, N.,
Jobes, F., Ono, Y., \& Perkins, F. 1997, Phys. Plasmas, 4, 1936.

Yokoyama, T. \& Shibata, K. 1994, ApJ Lett, 436, L197.

\clearpage

\clearpage

\clearpage

\clearpage

\begin{table*} [tbp]
\caption{Values of some basic plasma and reconnection region 
parameters in the typical coronal environment (column III) and 
in the flare environment (column IV). All values are calculated 
for pure hydrogen plasma ($Z=1$, $m_i=m_p$) and for $\zeta=2.0$ 
($\xi=e/\zeta \simeq 1.36$); $\tau_{\rm rec}$ is calculated for 
$L=10^9$ cm.}
\label{table}
\vskip 10 pt
\begin{center}
\begin{tabular*}{1.0\textwidth}{@{\hspace{2mm}}@{\extracolsep{\fill}}cccc}
\tableline \tableline
Parameter & Expression & Typical Coronal Value & Typical Flare Value \\ 
\hline

$T_e$, K   			& $T_e$				& 
$2\cdot 10^6$ 		&  $3.\cdot 10^7$ \\  
\tableline 

$T_i$, K   			& $T_i$				& 
$2\cdot 10^6$ 		&  $3.\cdot 10^6$ \\  
\tableline 

$n_e$, cm$^{-3}$  		& $n_e$				&  
$1 \cdot 10^9$ 		&  $1 \cdot 10^{10}$ \\  
\tableline 

$B_0$, G   			& $B_0$				& 
$100$ 			&  $100$ \\  
\tableline 

$v_s$, cm/sec			& $\sqrt{T_e/m_p}$		& 
$1.3\cdot 10^7$ &  $5\cdot 10^7$ \\  
\tableline 

$V_A$, cm/sec   		& $B_0/\sqrt{4\pi n_e m_p}$ 	& 
$6.9\cdot 10^8$ &  $2.2\cdot 10^8$ \\  
\tableline 

$\omega_{pe}$, sec$^{-1}$	& $\sqrt{4\pi n_e e^2/m_e}$	&
$1.8\cdot 10^9$ &  $5.6\cdot 10^9$ \\  
\tableline 

$\omega_{pi}$, sec$^{-1}$	& $\sqrt{4\pi n_e e^2/m_p}$	&
$4.2\cdot 10^7$ &  $1.3\cdot 10^8$ \\  
\tableline 

$\Omega_e$, sec$^{-1}$ 		& $eB_0/m_e c$			&
$1.8\cdot 10^9$ &  $1.8\cdot 10^9$ \\  
\tableline 

$\Omega_i$, sec$^{-1}$ 		& $ZeB_0/m_p c$			&
$9.6\cdot 10^5$ &  $9.6\cdot 10^5$ \\  
\tableline 

$\lambda_{De}$, cm   		& $\sqrt{T_e/4\pi n_e e^2}$	& 
$0.31$ 		&  $0.38$ \\  
\tableline 

$\lambda_{Di}$, cm   		& $\sqrt{T_i/4\pi n_e e^2}$	& 
$0.31$ 		&  $0.12$ \\  
\tableline 

$d_e$, cm 			& $c/\omega_{pe}$		&
$17$ 		&  $5.3$ \\  
\tableline 

$d_i$, cm 			& $c/\omega_{pi}$		&
$720$ 		&  $230$ \\  
\tableline 

$j_c$, cgs-units   		& $2.14\ en_e v_s$		& 
$1.3\cdot 10^7$ &  $5.1\cdot 10^8$ \\  
\tableline 

$\delta_c$, cm   		& $cB_0/4\pi j_c$		& 
$1.8\cdot 10^4$ &  $470$ \\  
\tableline 

$\eta_1$, cm$^2$/sec	& $1.4\cdot 10^{-3} (T_e/T_i) c^2/\omega_{pe}$& 
$7.1\cdot 10^8$	&  $2.2\cdot 10^9$ \\  
\tableline 

$\beta_e$    			&  $8\pi n_e T_e/B_0^2$		& 
$7.\cdot 10^{-4}$ 	&  $0.10$ \\  
\tableline 

$S_*$    		& $2\cdot 10^4 (T_i/T_e) V_A/c\sqrt{\beta_e}$	& 
$1.7\cdot 10^4$ 	&  $45$ \\  
\tableline 

$\Delta$, cm    		&  $\delta_c S_* \zeta^{-3}$	& 
$4\cdot 10^7$ 		&  $2.6\cdot 10^3$ \\  
\tableline 

$\tau_A(\Delta)$, sec   	&  $\Delta/V_A$			& 
$0.057$ 		&  $1.2\cdot 10^{-5}$ \\  
\tableline 

$V_{\rm rec}$, cm/sec   	& $V_A \zeta^2/S_*$		&
$1.6\cdot 10^5$		& $1.9\cdot 10^7$ \\
\tableline

$E$, cgs-units			& $4\pi \zeta^2 j_c\eta_1/c^2$	&
$5.2\cdot 10^{-4}$	& $6.4 \cdot 10^{-2}$ \\
\tableline

$\tau_{\rm rec}$, sec    	& $L/V_{\rm rec}$		&
$6.3\cdot 10^3$ 	& $52$ 	\\
\tableline

\tableline 
\end{tabular*}
\end{center}
\end{table*}

\clearpage

\end{document}